\documentclass[a4paper]{article}

\usepackage[english]{babel}
\usepackage[utf8x]{inputenc}
\usepackage[T1]{fontenc}

\usepackage[a4paper,top=3cm,bottom=3cm,left=3cm,right=3cm,marginparwidth=1.75cm]{geometry}

\usepackage{balance}  
\usepackage{graphics} 
\usepackage{txfonts}
\usepackage{times}    
\usepackage{color}
\usepackage{textcomp}
\usepackage{booktabs}
\usepackage{ccicons}
\usepackage{todonotes}
\usepackage[pdftex]{hyperref}
\usepackage{authblk}

\PassOptionsToPackage{pdfpagelabels=false}{hyperref}

\makeatletter
\def\mycopyrightnotice{%
  {\footnotesize This work is licensed under a Creative Commons Attribution License (CC BY 4.0).\hfill}
  \gdef\mycopyrightnotice{}
}
\def\url@leostyle{%
  \@ifundefined{selectfont}{\def\UrlFont{\sf}}{\def\UrlFont{\small\bf\ttfamily}}}
  \def\@oddfoot{\mycopyrightnotice}%
  \def\@evenfoot{\mycopyrightnotice}%
\makeatother
\def\pprw{8.5in}
\def\pprh{11in}

\definecolor{linkColor}{RGB}{6,125,233}
\hypersetup{%
  pdftitle={SIGCHI Conference Proceedings Format},
  pdfauthor={LaTeX},
  pdfkeywords={SIGCHI, proceedings, archival format},
  bookmarksnumbered,
  pdfstartview={FitH},
  colorlinks,
  citecolor=black,
  filecolor=black,
  linkcolor=black,
  urlcolor=linkColor,
  breaklinks=true,
}

\begin{document}

\title{User Longevity and Engagement in Mobile Multiplayer Sports Management Games}

\author[1]{Daniele Grassi}
\author[1]{Giacomo Barigazzi}
\author[2]{Giacomo Cabri}
\affil[1]{DM Digital, Via Rainusso 144, Modena, Italy}
\affil[2]{Universit\`a di Modena e Reggio Emilia, Via Campi, 213/B, Modena, Italy}


\maketitle

\begin{abstract}
  Mobile games are extremely popular and engage millions of people every day. Even if they are often quite simple, their development features a high degree of difficulty and requires close attention to both achieve a high satisfaction from users and grant a return to the developers. In this paper we propose a model that analyzes users' playing session time in order to evaluate and maximize the longevity of games, even during the first phases of their development.
\end{abstract}



\section{Introduction}

Ever since computers were introduced into our daily lives, digital games have radically transformed the way people spend their leisure time. New and accessible technology, such as mobile devices, and the possibility for developers to publish apps to public and centralized stores (like App Store, Play Store, Windows Phone Store) have created new opportunities for people to access videogames. The videogames industry has become the fastest growing leisure market, despite the recent global recession~\cite{chatfield2011fun}.

Like in any other business, the main focus of game developers concerns the quantity of consumers that use their product and their satisfaction. Game developers want to attract the highest number of users possible, and keep them engaged with their product for the longest time possible. Therefore, they need to understand what triggers consumers' engagement and leads them to spend time and money on the games created by the developers, and use this information as early as possible during the videogame's design and development process.

In this paper we present a study about user longevity and engagement in mobile games, and propose a model that analyzes the notion of daily playing time in order to evaluate and maximize games' longevity and engagement. We start from some hypothesis about the time users spend playing mobile games, and we build a model that connects different time parameters. Our model suggests how to size these parameters in order to maximize the total time spent playing the game. We do not address the issues related to how to improve players' engagement with the game in order to produce effects on their session time parameters, since this is very dependent on each single game. Our model can be used to validate the game design early during its development (providing a direct feedback) or to evaluate existing games.

In our work we have considered Multiplayer Sports Management Games (MSMGs), because they present the following features: a synchronous multiplayer online world that progresses even when the player is not actively engaged with the game, and a major game cycle (season) that spans across several days and weeks. Some examples of MSMGs are: Top Eleven\footnote{http://www.topeleven.com/}, Hattrick\footnote{https://www.hattrick.org/}, Online Soccer Manager\footnote{http://en.onlinesoccermanager.com/}, GOAL 15: Be A Football Manager\footnote{http://www.goal-games.de/en/home}, Top Race Manager\footnote{http://topracemanager.com/}.

The rest of the paper is organized as follows. The next section
reports on related work. Then, we will 
introduce the hypothesis we have taken in our work and the parameters we exploit in our model. 
After that, we will 
present the proposed model to address longevity and engagement in mobile games. 
Finally, we will 
conclude the paper and sketch some future work.

\section{Related Work}
\label{sec:RelatedWork}

Existing literature has tried to define a set of tools and features that are similar in successful games. This issue has been approached both from a social/psychological and a technical point of view: with respect to the former, the main issue concerns the research of the psychological elements that make a game enjoyable; from the technical point of view, the goal is to develop a game with a set of features as determined by such psychological elements. According to Connolly et al.~\cite{stansfield2004enhancing}, ``computer games build on theories of motivation, constructivism, situated learning, cognitive apprenticeship, problem-based learning, and learning by doing''. Griffiths \& Davies~\cite{griffiths2002research} suggest that computer games incorporate features that have compelling, or even addictive, qualities.

A deep analysis of motivation was performed by Deci \& Ryan~\cite{ryan1991motivational}, showing a distinction between intrinsic and extrinsic motivation. Intrinsically motivated behaviors are connected with self-rewarding while extrinsically motivated behaviors are usually triggered by the desire for some external reward, such as money, praise or recognition from others.

Malone and Lepper~\cite{malone1987making} argued that intrinsic motivation prevails in designing engaging games and suggested that intrinsic motivation is driven by four individual factors-challenge, fantasy, curiosity and control—, as well as three interpersonal factors-cooperation, competition, and recognition.
Extrinsically motivated behaviors are extremely important in managerial multi-player games, as the user plays against other users, not machines. Users that already know each other (e.g. through connections on social networks) often play together or against one another in multi-player games and this increases the extrinsic motivation. Gajadhar, W. de Kort and Iksselsteijn~\cite{gajadhar2009rules} performed an experiment involving 86 players and found that playing with others contributed to the players' involvement in the game. They also concluded that player involvement is not necessarily impaired by the presence of others.
To achieve the goal of higher player involvement, our investigation focused on how games were designed to be enjoyable, what elements appealed to the players and what encouraged them to continue playing. According to Prensky~\cite{prensky2001digital}, technology has allowed a more immersive game-play experience thanks to the introduction of augmented reality gaming that extends learning and enjoyment when the natural scene of the game is enhanced by the addition of digital objects.

S. de Freitas~\cite{de2007serious} suggests that two different methods can be used to increase player involvement and enjoyment. The first method is to ensure that the player has the feeling of being inside the game, while being able to play with other participants, either competing or collaborating.
The second is to provide a clearly articulated set of goals and sub-goals, such as completing a level, obtaining intermediate awards, etc. In managerial games, short term, mid-term and long term goals must be connected with a set of rewards that reinforces players' engagement and self-awareness~\cite{kambouri2003designing}.

Thus, although several authors stress the importance of psychological elements such as rewards and social interaction, there has been little concrete research on the practical aspects of the industry, for instance consumer retention. Most works and literature in this regard have predominantly addressed psychological concepts and theories. We haven't found any academic papers that specifically address the estimate of the time that users spend on games, or how to increase the time and money spent by them in the game, or about frameworks that allow a simple validation of games design early in their development process. Considering the current market growth though, we may well conclude that this will change soon.

\section{Model Hypothesis}
\label{sec:ModelHypothesis}

Before presenting our model, in this section we define some hypothesis and terms taken as foundation of the model.

\subsection{Limited player's daily playing time}
\label{sec:LimitedPlayerSDailyPlayingTime}

Mobile gaming is characterized by short playing sessions that are performed in a non-structured schedule during the day, with an average playing time of 30-45 minutes per day~\cite{FlurryIV}; this time is divided by each player among the different games he plays on a semi-regular basis. Our assumption is that a player would dedicate at least half of his daily playing time to the game he prefers the most at that time (during a player's lifetime, the set of games he plays on a daily basis changes several times). This implies that, when designing a mobile game, the designer should be aware that the daily time that even the most engaged player can dedicate to it is limited on average. ~\cite{FlurryIV} shows how the average daily time spent on games varies among countries.

\subsection{Limited player's total playing time per game}
\label{sec:LimitedPlayerSTotalPlayingTimePerGame}

Multiplayer sports management games (MSMGs) usually span their main playing cycle (represented by the in-game virtual season) over several days or weeks; successful MSMGs retain players over several in-game seasons, during which the players can experience progress and achievements in the in-game world. In general, games based on an online in-game synchronous world (such as MSMGs) feature an inviolable upper limit on how much the player can progress each day: the horizon of the game becomes thus several weeks or months.

These characteristics make MSMGs (and, in general, games based on an online in-game synchronous world) exhibit very different engagement dynamics from the ones featured by games where the players' achievements and progress speed depend exclusively on the players' playing time and will. An example of the latter is represented by games where a player doesn't experience a ``daily upper limit'' to his progress, and they can play it continuously completing levels and missions until they want to. Nevertheless, it should be noted that many games that fall into this category use techniques to limit players daily activity for commercial reasons, making players purchase additional in-game credits to keep playing after a certain limit or time each day. 

MSMGs usually present both synchronous and asynchronous aspects. They feature a synchronous in-game world with scheduled matches and season events: this allows players to prepare for matches where their teams play against each other. Players activities consist mainly of asynchronous tasks and decisions that affect the behaviour and results of their teams (e.g. asynchronously setting a lineup for a match that will be played at a scheduled time by the game engine). Usually, as long as a task is performed before a certain synchronous in-game event, the specific time on which the task is completed doesn't matter. 
The synchronous in-game world requires the user to extend his playing time over several days, weeks and months to achieve results. It's clear how these dynamics require a certain commitment from the player, a commitment that features a different ``deepness'' than the commitment required by a truly casual game where a player-independent schedule is not present. Additionally, technology, mobile devices and their capabilities change over time, and a large number of new mobile games are published daily (in 2014, an average of 1,385 games were submitted to the Apple App Store per day~\cite{Pocket}), pushing players to try new ones every day: older games tend to seem thus ``obsolete'' and to be substituted by players in their ``currently playing games'' set.

All this said, we assume that, on average, the total time that a player can dedicate during his lifetime to a specific MSMG has an upper limit, and that it is possible to determine an average of total player's playing time per MSMG ($T_M$; i.e. the total time a player will play—during his lifetime—to a particular game). This time $T_M$ is equal to the average total playing time that designers can expect from players given a specific game type; we can thus expect that a player, on average, given a sufficient level of engagement provided by a game, will play with it only up to a total time in the long-term. This assumption is considered valid, for the scope of this paper, only for MSMGs (although similar assumptions can be made for games that feature a synchronous in-game world and for games that actively limit the daily free-play, further study should be done to assess these assumptions). 

A method to determine $T_M$ could be a survey delivered to MSMG players asking for the frequency, the average daily session and the total period they played with each MSMG they liked but that they don't play with anymore. A qualitative estimate we provide is 15 minutes each day (calculated by taking half of the average time that players dedicate to games daily) for 6 months (the time after which commitment to a specific game appears to fade), for a total time of around 45 hours.

It can be argued that the ``sufficient level of engagement provided by a game'' is a value that cannot be measured, and that the very significance of $T_M$, being an average, depends on its standard deviation being sufficiently low: further study should be performed on the matter. For the time being, we take as reference the HowLongToBeat website\footnote{http://howlongtobeat.com} that provides user-generated statistics about games lifespan: analyzing mobile games listed on the site, it appears that average games' total playing time (mostly arcade and adventure) should be between 15 and 25 hours. Our estimation of 45 hours for MSMGs seems plausible in this respect since MSMGs require a long-term commitment (and shorter daily sessions) that should improve the games' lifespan.

\subsection{Player's daily critical playing time}
\label{sec:PlayerSDailyCriticalPlayingTime}

MSMGs, and in general games that feature synchronous in-game worlds, usually provide the players with a daily set of tasks that the player should perform to optimize their progress in the game (for example, checking training results, setting up the formation for the next match, etc.). These tasks, although they don't specifically and individually determine the long-term results of the player, should provide the player with a feeling of improvement and progress in short-term goals (such as winning a single match) and short-term achievements (e.g. reaching a 3-match winning streak); they should also keep the player busy for enough time to let them feel ``immersed'' in the in-game world, letting them tune their thoughts with it and thus being in a mental state that allows them to feel engaged and rewarded by short-term improvement and progress. We define $t_c$ as the average minimum daily playing time necessary for the player to experience sufficient reward from the game and to develop an affection for it.
$t_c$ depends on the game design, but we qualitatively find that MSMGs (or, in general, games featuring similar synchronous in-game world's paces)  have similar $t_c$; $t_c$ can be seen also as the lower limit for daily playing time under which the game cannot make its way in the daily routine of the player, being thus quickly forgotten.

In our experience, in MSMGs $t_c$ is between 3 and 5 minutes (further study should be performed to get a clearer measure of $t_c$).

\subsection{Daily playing time maximizing total player's playing time}
\label{sec:DailyPlayingTimeMaximizingTotalPlayerSPlayingTime}

Given the previous assumptions, game designers should aim to maximize the total player's playing time for their game by carefully engineering player's rewards and experience of progress in a way that syncs with the intrinsic in-game cycle (avoiding providing reward and progress experience ``too rarely'' or ``too often'' compared with the pace of the in-game world). This means, in fact, designing the game to provide ``a sufficient level of engagement'' during the time that, given the game's type, the player concedes to the game itself.

We define the daily playing time maximizing total player's playing time ($t_M$) as the average daily session time that allows the game to provide an optimum amount of reward to the player, maximizing $T_M$. In other words, $t_M$ is the daily playing time that allows the player to get a firm grasp of their progress and to feel sufficient engagement with the in-game world to maximize their long-term commitment to the game.

\section{The Proposed Model}
\label{sec:TheProposedModel}

\subsection{Game's required daily playing time}
\label{sec:GameSRequiredDailyPlayingTime}

The game's required daily playing time $t_r$ strictly depends on the specific game's design, and identifies the average session time a player should spend on the game every day to achieve at least a slow but perceivable progress and rewards (both in terms of short-term improvement and of long-term results).

For example, the $t_r$ of an arcade game could be the average daily session time needed for the player to improve their performance day after day. In a MSMG, $t_r$ would represent the average daily session time needed to improve a player's standings and level across the in-game seasons.

In the specific case of mobile online management games, where the game's internal calendar is mostly fixed and players execute mostly asynchronous operations between them, $t_r$ directly depends on what operations are available for the player each time they log in the game, and on the impact those operations have on their performance. These activities don't necessarily have to obtain an immediate impact on the player's results, but should at least provide the player with a short-term feeling of improvement and show a clear connection between them and the improved possibility to reach long-term results.

In general, playing with a game for less than its $t_r$ would lead to poor results and almost no progress in the game.

\subsection{Game's ideal daily playing time}
\label{sec:GameSIdealDailyPlayingTime}

Depending on its design and mechanics, every game has an ideal daily playing time ($t_i$). We define $t_i$ as the average daily time a player has to spend on the game to obtain optimal in-game long-term progress and rewards; in other words, it's the average daily time a player should spend to reach a point where adding more daily playing time to the same game wouldn't sensibly improve their long-term performance further.

\subsection{Average players' results as a function of daily session time}
\label{sec:AveragePlayersResultsAsAFunctionOfDailySessionTime}

\begin{figure}
  \centering
  \includegraphics[scale=0.90]{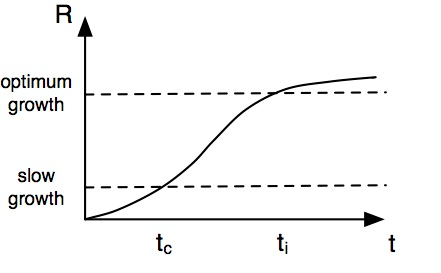}
  \caption{Relationship between the game time $t$ and the in-game results $R$.}
	\label{fig:figure1}
\end{figure}

Assuming an average innate gaming skill for players, we can relate players' in-game results (R) with their daily session average playing time ($t$).  As per $t_i$ definition, the ``optimum growth'' result level intersects the results curve in correspondence with $t_i$ (see Figure~\ref{fig:figure1}).

Deriving from $t_c$ definition, we can position it in correspondence with the intersection of the results curve and of the slow growth results level (being this the lowest growth speed providing the player with sufficient progress and rewards to engage them and make them commit to the game).

\subsection{Designing game dynamics to optimize total players' playing time}
\label{sec:DesigningGameDynamicsToOptimizeTotalPlayersPlayingTime}

As described before, in this framework game designers have three variables they can use to tweak their game and players' engagement: $t_r$, $t_i$  and  $t_M$.
It's important that tr is higher than $t_c$, so that the minimum average session time required to obtain progress in the game is more than the critical time: if it were lower, players who find themselves playing with the game strictly for the time necessary to ``survive'' in it (for example in periods of little free time, or simply of low temporary engagement with the game) would disengage very quickly and abandon the game.

At the same time, the steepness of the reward curve between $t_r$ and $t_i$ should be high: this way, players would be very encouraged to spend more time in the game because a little difference in effort would mean a bigger difference in results.

$t_M$, which is the average daily session time that maximises overall game longevity (in terms of total played hours), should be positioned between $t_r$ and $t_i$: this way, it would be easy to reach (being in a section where every additional playing minute to the game rewards the player significantly).

\begin{figure}
  \centering
  \includegraphics[scale=0.90]{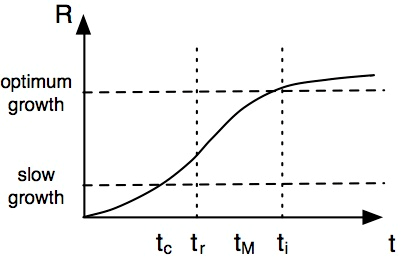}
  \caption{Distribution of the different times.}
	\label{fig:figure2}
\end{figure}

The distribution of $t_r$, $t_i$  and  $t_M$ is reported in Figure~\ref{fig:figure2}.

\section{Conclusions}
\label{sec:Conclusions}

In this paper we have proposed a model to evaluate MSMGs and model the different playing time parameters that characterize it in order to maximize the longevity and the players' engagement. We have considered Multiplayer Sports Management Games (MSMGs) because they exhibit features and dynamics that require players to play consistently across several days and weeks to experience progress, and because they feature synchronous in-game world that progress independently from the single player interaction with the game. Designing MSMGs to obtain time parameters as suggested by our model will help game developers to achieve the appropriate involvement from the users, increasing the game's longevity and cumulative playing time (which usually correlates with economic return for the game developers). 

Future work will deepen the study of specific research topics. First, we will evaluate the standard deviation of $T_M$ (i.e. the total time a player will play a specific game during his lifetime) in order to understand how this value can be significant. Second, we will study a clearer definition of $t_c$, i.e. the average minimum daily playing time necessary for the player to experience sufficient reward from the game and to develop an affection for it. Third, even if MSMG games are very representative of several kinds of game, we will evaluate the applicability of our framework to other kinds of games.

\vspace{20pt}

\bibliographystyle{plain}
\bibliography{chiplay15}

\begin{thebibliography}{10}

\bibitem{chatfield2011fun}
Tom Chatfield.
\newblock {\em Fun inc: why games are the twenty-first century's most serious
  business}.
\newblock Random House, New York, NY, USA, 2011.

\bibitem{de2007serious}
Sara De~Freitas and Steve Jarvis.
\newblock Serious games-engaging training solutions: A research and development
  project for supporting training needs.
\newblock {\em British Journal of Educational Technology}, 38(3):523, 2007.

\bibitem{FlurryIV}
Jarah Euston.
\newblock Gaming: The lingua franca of mobile, 2014.

\bibitem{gajadhar2009rules}
Brian~J Gajadhar, YAW de~Kort, and Wijnand~A IJsselsteijn.
\newblock Rules of engagement: Influence of co-player presence on player
  involvement in digital games.
\newblock {\em International Journal of Gaming and Computer-Mediated
  Simulations (IJGCMS)}, 1(3):14--27, 2009.

\bibitem{Pocket}
Pocket Gamer.
\newblock App store metrics, 2015.

\bibitem{griffiths2002research}
M.D. Griffiths and M.N.O. Davies.
\newblock Excessive online computer gaming: implications for education.
\newblock {\em Journal of Computer Assisted Learning}, 18(3):379--380, 2002.

\bibitem{kambouri2003designing}
Maria Kambouri, Siobhan Thomas, and Gareth Schott.
\newblock Designing for learning or designing for fun? setting usability
  guidelines for mobile educational games.
\newblock {\em Proceedings of MLEARN. Retrieved on March}, 23:2009, 2003.

\bibitem{malone1987making}
Thomas~W Malone and Mark~R Lepper.
\newblock Making learning fun: A taxonomy of intrinsic motivations for
  learning.
\newblock {\em Aptitude, learning, and instruction}, 3(1987):223--253, 1987.

\bibitem{prensky2001digital}
Marc Prensky.
\newblock Digital natives, digital immigrants part 1.
\newblock {\em On the horizon}, 9(5):1--6, 2001.

\bibitem{ryan1991motivational}
Richard~M Ryan.
\newblock A motivational approach to self: Integration in personality edward
  l., deci and.
\newblock {\em Perspectives on motivation}, 38:237, 1991.

\bibitem{stansfield2004enhancing}
Mark Stansfield, Evelyn McLellan, and Thomas Connolly.
\newblock Enhancing student performance in online learning and traditional
  face-to-face class delivery.
\newblock {\em Journal of Information Technology Education: Research},
  3(1):173--188, 2004.

\end{thebibliography}

\end{document}